\documentclass[a4paper,twocolumn,english,showpacs,aps,prr,floatfix,superscriptaddress]{revtex4-2}
\usepackage[utf8]{inputenc}
\usepackage{amsmath}
\usepackage{amsthm}
\usepackage{amssymb}
\usepackage{graphicx}
\usepackage[caption=false]{subfig}
\usepackage{xparse}
\usepackage{xcolor}
\usepackage{siunitx}
\usepackage{hyperref}

\usepackage{float}

\usepackage{booktabs}
\usepackage{tabularx}

\usepackage{multirow}

\usepackage{times}
\makeatletter

\newcounter{thmc}

\newtheorem{lemma}[thmc]{Lemma}

\usepackage{braket}
\usepackage{dsfont}

\def\>{\rangle}\def\<{\langle}

\def\section#1{{\em #1:}}

\makeatother

\usepackage{babel}

\global\long\def\ketbra#1{\ket{#1}\!\bra{#1}}

\newcommand{\kommentar}[1]{}

\NewDocumentCommand\opti{smmm>{\SplitList{;}}m} {
\begingroup%
\setlength{\belowdisplayskip}{-0.6\baselineskip}%
\IfBooleanTF{#1}{%
    \begin{alignat*}{2}
        & \underset{#3}{\text{#2}} & & #4 \\
        & \text{subject to~~}
        \ProcessList{#5}{ \insertopticonst }
        & &
    \end{alignat*}%
    }{%
    \begin{alignat}{2}
        & \underset{#3}{\text{#2}} & & #4 \\
        & \text{subject to~~}
        \ProcessList{#5}{ \insertopticonst }
        & & \nonumber
    \end{alignat}%
    }%
\endgroup%
}%
\newcommand\insertopticonst[1]{& & #1\\&}

\begin{document}

\title{Complementarity-based complementarity: the choice of mutually unbiased observables shapes quantum uncertainty relations}

\author{Laura Serino}
\affiliation{Paderborn University, Integrated Quantum Optics, Institute for Photonic Quantum Systems (PhoQS), Warburgerstr.\ 100, 33098 Paderborn, Germany}

\author{Giovanni Chesi}
\affiliation{QUIT Group, Physics Department, Univ. Pavia, INFN Sez.~Pavia, Via Bassi 6, 27100 Pavia, Italy}

\author{Benjamin Brecht}
\affiliation{Paderborn University, Integrated Quantum Optics, Institute for Photonic Quantum Systems (PhoQS), Warburgerstr.\ 100, 33098 Paderborn, Germany}

\author{Lorenzo Maccone}
\affiliation{QUIT Group, Physics Department, Univ. Pavia, INFN Sez.~Pavia, Via Bassi 6, 27100 Pavia, Italy}

\author{Chiara Macchiavello}
\affiliation{QUIT Group, Physics Department, Univ. Pavia, INFN Sez.~Pavia, Via Bassi 6, 27100 Pavia, Italy}

\author{Christine Silberhorn}
\affiliation{Paderborn University, Integrated Quantum Optics, Institute for Photonic Quantum Systems (PhoQS), Warburgerstr.\ 100, 33098 Paderborn, Germany}

\date{\today}

\begin{abstract}
Quantum uncertainty relations impose fundamental limits on the joint knowledge that can be acquired from complementary observables: perfect knowledge of a quantum state in one basis implies maximal indetermination in all other mutually unbiased bases (MUBs). Uncertainty relations derived from joint properties of the MUBs are generally assumed to be uniform, irrespective of the specific observables chosen within a set. In this work, we demonstrate instead that the uncertainty relations can depend on the choice of observables. Through both experimental observation and numerical methods, we show that selecting different sets of three MUBs in a 5-dimensional quantum system results in distinct uncertainty bounds, i.e. in varying degrees of complementarity, in terms of both entropy and variance. 
  \typeout{We show that the amount of complementarity that a quantum
    system can exhibit depends on which complementary properties one
    is considering. Complementary properties can be connected to
    mutually unbiased bases (MUBs): if the value of one property is
    known (i.e. the system state $|\psi\>$ is in one of the basis
    states), then the complementary properties, described by bases
    that are mutually unbiased to the basis to which $|\psi\>$
    belongs, are completely unknown: the measurement of another
    property will find any of its possible outcomes with uniform
    probability. Here we show that a 5-dimensional system can have
    different degrees of complementarity, depending on which three of
    the six MUBs we choose. The degree of complementarity is assessed
    using uncertainty relations, entropic or variance
    based. Interestingly, this result was first found experimentally,
    in the experiment detailed here.}
\end{abstract}
\maketitle

A property of a quantum system is described by an observable, namely a
Hermitian operator $\hat{O}$. Each of the possible values $o$ of the
property $\hat{O}$ is connected to an eigenstate $|o\rangle$. The Born-rule
probability that the measurement of a system property has outcome
equal to the value $o$ is given by $|\langle o|\psi\rangle|^2$, where
$|\psi\rangle$ is the system state. This state of affairs formalizes
the principle of complementarity \cite{bohr1928,bohrbohr1928}: A
system can possess a definite value of a property, i.e. measurement
outcomes have probability 1, only if its state is an eigenstate
$|\psi\rangle=|o\rangle$ of that property. Otherwise, the value of
that property is undefined, and only probabilistic predictions of
measurement outcomes are possible. This implies that complementary properties
exist: From a set of properties of the system, in general we can assign a
definite (i.e.~fully determined) value only to one.

In particular, maximally complementary properties exist: Assigning a definite value to one of them renders {\em all} the others maximally indeterminate, namely each of their outcomes will have uniform probability. This happens when we consider a set of observables whose eigenstates are mutually unbiased bases (MUBs). Indeed, the square modulus of the scalar product of any 
two states $|a_i\rangle, |b_j\rangle$ pertaining to two different MUBs is always $|\langle a_i|b_j\rangle|^2=1/d$, where $d$ is the Hilbert space dimension.

Surprisingly, it was shown \cite{wootters1979, greenberger1988, englert1996} that one can have large information on more than one complementary 
observable provided that the value of none of them is known with certainty (which would render the values of all the others completely 
indetermined). Namely, if the system is prepared in a state that is a nontrivial superposition when expressed in all the MUBs, then it is 
possible to have nontrivial {\em joint} information on multiple complementary properties.

\begin{figure} [htb]
    \includegraphics[width=0.38\textwidth]{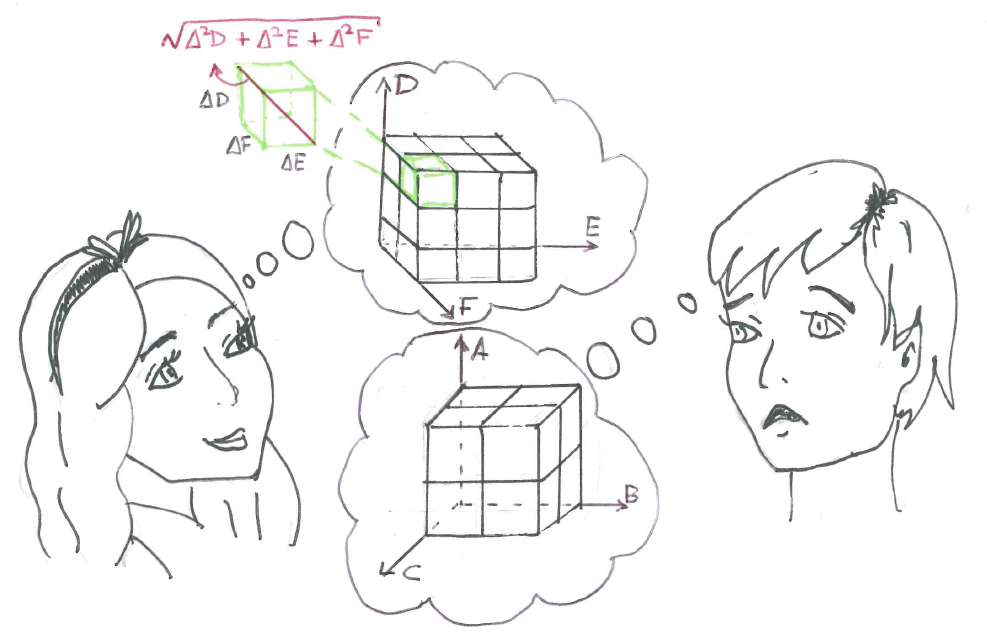}
    \caption{
      Alice and Bob are both interested in joint values of
      observables with MUBs as eigenstates. In
      dimension 5 there are 6 of them: Bob measures
      observables $\hat{A},\hat{B},\hat{C}$, Alice measures $\hat{D},\hat{E},\hat{F}$. 
      Alice gets more information: she can divide the $DEF$ phase space in smaller uncertainty blocks than Bob's, even though they are both looking at MUBs.
      }
    \label{f:alice}
\end{figure}

Different complementary properties (MUBs) can be grouped and classified in terms of equivalence classes of complex Hadamard matrices
\cite{tadej2006,durt2010,brierley2010}. We say that two sets of MUBs are equivalent if one can be mapped into the other by unitary transformations, permutations and phase factor multiplications. In this context, it was pointed out in Ref.~\cite{brierley2010} that in dimension $d=5$ there are two \textit{inequivalent} classes of triplets of MUBs. For larger dimensions, a rich and involved structure of inequivalent sets of MUBs is known to exist \cite{durt2010,kantor2012,sehrawat2014}. In Ref.~\cite{designolle2019}, a notion of \textit{operational inequivalence} of sets of MUBs has been introduced, such that the inequivalence is detected whenever distinct sets of MUBs feature a different noise robustness, used as an incompatibility quantifier. 
\\
The existence of inequivalent sets of MUBs was shown to have practical consequences on Quantum Random Access Codes, since it prevents to generalize the analytic expression of the optimal average success probability for protocols $2^d \to 1$ to protocols $n^d \to 1$ with $n\geq 3$ \cite{aguilar2018}. 
Applications have also been found in entanglement detection based on complementary observables: in Ref.~\cite{hiesmayr2021} it was shown that the lower bound on a given correlation function, used as an entanglement witness, depends not only on the dimension and the number of measurements, but also on the choice of the specific set of MUBs.
\\
While these contributions were mainly focused on proving and/or detecting the existence of inequivalent sets of MUBs, here we present a fundamental implication of this: their impact on the uncertainty relations (URs). The URs are the very expression of the complementary properties of quantum observables and yield a direct link between experimental observations and quantum theory. While providing experimental validation of the URs for MUBs and high-dimensional systems, we detected different minimal uncertainties for distinct triplets of MUBs in $d=5$. 
Then, the inequivalence proved in Ref.~\cite{brierley2010} unvealed a finer structure of quantum URs: The information that a system can possess on three of the six complementary properties depends on {\em which} of the properties are considered. We show this by calculating the lower bound of the sum of the entropies or of the variances of the measurement outcomes of three complementary observables, complemented with Monte-Carlo simulations, and prove that such a lower bound depends explicitly on the choice of complementary observables:
Complementarity-based complementarity (Fig.~\ref{f:alice}). Most importantly,
the validity of distinct URs for triplets in the same complete set of MUBs opens foundational questions,
finds application in a plethora of quantum protocols and 
yields to experimental verification, which we report here.
%


\section{
Uncertainty relations for inequivalent triplets of MUBs}
\label{2}
Consider three maximally complementary observables $\hat{A},\hat{B},\hat{C}$ with mutually-unbiased
eigenstates $\{|a_j\>\}_j,\{|b_j\>\}_j,\{|c_j\>\}_j$.
One can have {\em partial} knowledge of all three of them if the state
$|\psi\>$ is a nontrivial superposition when expressed in any of the
bases $\{|a_j\>\}_j,\{|b_j\>\}_j,\{|c_j\>\}_j$. How much {\em joint} information on them
can one obtain? We need a quantification of how uncertain the outcomes
of measurements of all three observables are. This is precisely what we learn from URs. In the following, we concentrate on two quantifiers \footnote{We also considered the Renyi entropies of order 2 and 1/2, but the numerics gave inconclusive results}. On the one
hand, we consider the sum of the Shannon entropies $H(\hat{O}) = -\sum_{j}p_j\log_2p_j$ of the Born
probabilities  $p_j = |\langle o_j|\psi\rangle|^2$ pertaining to the measurement of $\hat{O}$.
The entropies are all positive quantities, so a small
sum implies large {\em joint} knowledge. 
The sum of entropies for maximally complementary observables has always a
nontrivial lower bound given by the entropic uncertainty relations
(EURs) \cite{deutsch1983,maassen1988,riccardi2017}. On the other hand, we
consider the sum of the variances \cite{trifonov1998} of $\hat{A},\hat{B},\hat{C}$. Differently from the entropy, the variance depends
also on the eigenvalues of the observables, not only on the probabilities. We will choose observables with eigenvalues equal
to a permutation $P(j)$ of the basis index $i$, i.e.~$\hat{A}=\sum_jP(j)|a_j\>\<a_j|$, and then minimize the variance $\Delta A^2=\<\hat{A}^2\>-\<\hat{A}\>^2$ over the permutations, to avoid effects due to the arbitrariness of the eigenvalue assignments. Also the sum of
variances has a nonzero \footnote{Instead, the product of variances, as in the
traditional Heisenberg-Robertson uncertainty relations \cite{robertson1929}, is problematic because it has no nonzero lower
bound when minimizing over states: the product is null already if only one of the two variances is nonzero.} lower bound for maximally complementary
observables \cite{trifonov2000,trifonov2002}. 
We now show that the 
UR in terms of entropies or variances of three different maximally complementary observables $\hat{A},\hat{B},\hat{C}$ depends on {\em
  the choice of the maximally complementary observables}.

We start from dimension $d=5$ where there exist six maximally complementary
observables $\hat{A},\hat{B},\hat{C},\hat{D},\hat{E},\hat{F}$ with eigenvectors equal to the six MUBs \cite{brierley2010,durt2010} 
presented in the appendix.  In this case, we find that
\begin{align} \label{euralberto}
  &H(\hat{A}) + H(\hat{B}) + H(\hat{C}) \geq 2\log_25\simeq 4.64386\\
  &H(\hat{D}) + H(\hat{E}) + H(\hat{F}) \gtrsim 4.43223.
 \label{eursurprise}
\end{align}
The bound \eqref{euralberto} was known~\cite{riccardi2017} and was implicitly assumed to hold for every possible choice of MUB triplet, whereas the bound \eqref{eursurprise}, obtained numerically (see in the appendix), was found while experimentally testing the EURs. The states that minimize the first bound are 
eigenstates of any of the three $\hat{A},\hat{B},\hat{C}$, e.g.~$|\psi\>=|a_0\>$. The
states that minimize the second bound, instead, have a
single null component when expressed in the computational basis
$|a_j\>$, e.g.~a state $\sum_j\psi_je^{i\phi_j}|a_j\>$ with
$\{\psi_j\}=(0.54488, 0.45067, 0, 0.45067, 0.54488)$ and
$\{\phi_j\}=\left(0, -{2}\pi/{5}, 0, \pi, -{\pi}/{5}\right)$.  

\begin{figure} [htb]
    \includegraphics[width=0.48\textwidth]{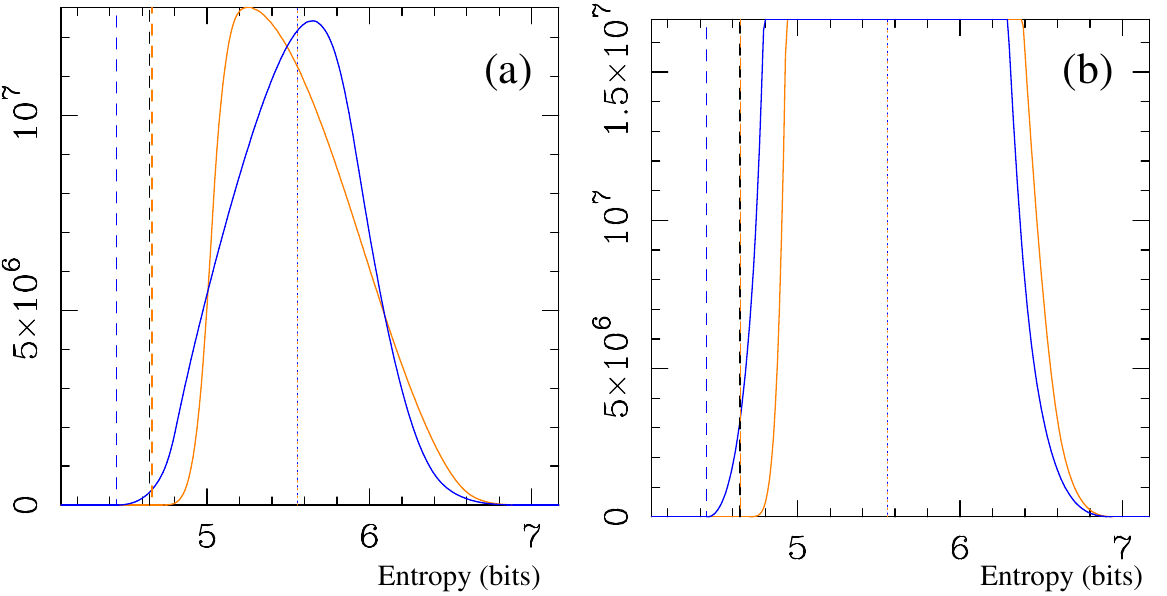}
    \caption{Monte-Carlo evaluation of the sums of entropies in
      \eqref{euralberto} and \eqref{eursurprise} on pure states chosen
      randomly using the Haar measure. The histograms (250 bins)
      represent the number of states whose sum of entropies of three
      maximally complementary observables are equal to the value in
      the abscissa. (a)~Simulation over $10^9$ random states. (b)~Detail of the tails of the distributions for a simulation over $10^{10}$ random
      states. The orange (blue) curves refer to the entropies of
      $A,B,C$ ($D,E,F$). The black dashed line is at $2\log_25$ and is
      approached by the left tail of the orange distribution, the blue
      dashed line approaches the lower bound of
      \eqref{eursurprise}. The dotted lines are the (matching) average
      values of the two distributions at $\sim 5.55$.}
    \label{f:montecarlo}
\end{figure}

\begin{figure} [htb]
    \includegraphics[width=0.48\textwidth]{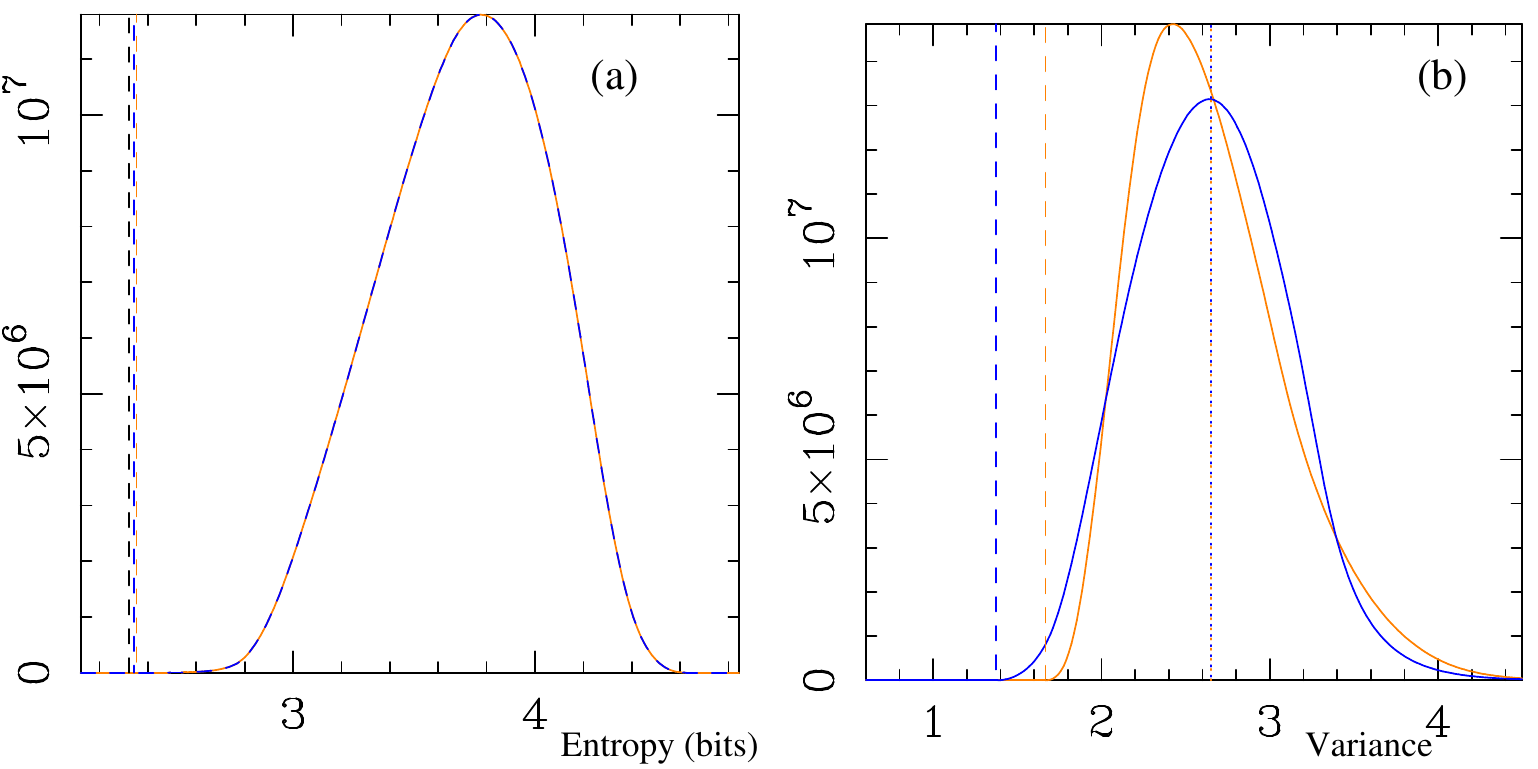}
    \caption{(a)~Histograms of the Monte-Carlo evaluation of the sum
      of the entropies of two MUBs in dimension $d=5$ (250 bins over
      $10^9$ Haar distributed random pure states). As expected, all
      histograms match (here the histograms refer to $A,B$ in orange
      and $C,D$ in blue). The lower bounds of the left tails
      (orange-blue vertical dashed lines) approach the Maassen-Uffink
      bound $\log_25$ (black dashed line). (b)~Histograms (250 bins
      over $10^9$ states) of the sum of variances of three MUBs, evaluated over $A,B,C$ (orange) with
      lower bound 1.67, and $D,E,F$ (blue) with bound 1.37.
      }
    \label{f:mu}
\end{figure}

To
validate the numerical minimization, we calculated the minimal sum of the entropies on a large number of pure states chosen randomly
\cite{zyczkowski1994,zyczkowski1998} with Haar measure (Fig.~\ref{f:montecarlo}), confirming that no state
beats either bound \eqref{euralberto}, \eqref{eursurprise}. We also
checked that the sum of any {\em two} entropies never beats the
Maassen-Uffink bound \cite{maassen1988} of $\log_25$ (Fig.~\ref{f:mu}) and that the sum of $four$ entropies has the unique lower bound found in Ref.~\cite{riccardi2017} for every choice of the MUBs,
confirming that the effect presented here requires exactly three
MUBs.

We inspected all the $\binom{6}{3}=20$ possible combinations of three
MUBs out of 6 and found only the two bounds reported in Eqs.~\eqref{euralberto} and~\eqref{eursurprise}. Moreover, if a triplet could
attain one bound, the remaining triplet would obtain the other
bound. In particular, the triplets that only achieve the bound
$2\log_25$ are the ones in the set 
\begin{align} \label{s1}
\mathcal{S}_1 \equiv \{&ABC,ABF,BEF,ADE,BCD,CEF,CDF,BDE, \nonumber \\
&ACE,ADF\}; 
\end{align}
whereas the ones that achieve the bound in Eq.~\eqref{eursurprise} are the triplets in the set 
\begin{align} \label{s2}
\mathcal{S}_2\equiv \{&ABD,ABE,ACF,BCE,DEF,CDE,BDF,ACD,\nonumber \\
&BCF,AEF\}.
\end{align}
\begin{center}
\begin{table} 
\begin{tabular}{ |c|c|c|c|c|c| }
\hline
$\{l_1,l_2,l_3\}$ & $\psi_0$ & $\psi_1$; $\phi_1$ &$\psi_2$; $\phi_2$ & $\psi_3$; $\phi_3$ & $\psi_4$; $\phi_4$ \\
\hline
\hline
\multirow{2}{2.5em}{$ABD$} & \multirow{2}{2em}{0.23} & 0.67 & 0.67 & 0.23 &  0 \\ 
& & $4\pi/5$ & $-\pi/5$ & $\pi$ & 0 \\
\hline
\multirow{2}{2.5em}{$ABE$} & \multirow{2}{2em}{0.23}  & 0 & 0.23 & 0.67 & 0.67 \\
&  & 0 & $\pi$ & $\pi/5$ & $-4\pi/5$ \\
\hline
\multirow{2}{2.5em}{$ACD$} & \multirow{2}{1em}{0} & 0.23 & 0.67 & 0.67 & 0.23 \\
&  & 0 & $4\pi/5$ & $\pi/5$ & $\pi/5$ \\
\hline
\multirow{2}{2.5em}{$ACF$} & \multirow{2}{2em}{0.67}  & 0.23 & 0.23 & 0.67 & 0 \\
&  & $-3\pi/5$ & $4\pi/5$ & $\pi/5$ & 0 \\
\hline
\multirow{2}{2.5em}{$AEF$} & \multirow{2}{2em}{0.23}  & 0.67 & 0.67 & 0.23 & 0 \\
&  & $2\pi/5$  & $\pi/5$ & $-3\pi/5$ & 0 \\
\hline
\multirow{2}{2.5em}{$BCE$} & \multirow{2}{2em}{0.45}  & 0.45 & 0.55 & 0 & 0.55 \\
&  & $-\pi/5$ & $2\pi/5$ & 0 & $\pi$ \\
\hline
\multirow{2}{2.5em}{$BCF$} & \multirow{2}{2em}{0.45}  & 0.45 & 0.55 & 0 & 0.55 \\
&  & $-3\pi/5$ & $4\pi/5$ & 0 & $3\pi/5$ \\
\hline
\multirow{2}{2.5em}{$BDF$} & \multirow{2}{2em}{0.45}  & 0.45 & 0.55 & 0 & 0.55 \\
&  & $\pi$ & $-4\pi/5$ & 0 & $\pi/5$ \\
\hline
\multirow{2}{2.5em}{$CDE$} & \multirow{2}{2em}{0.55}  & 0.55 & 0.45 & 0 & 0.45 \\
&  & $-\pi/5$  & $\pi/5$ & 0 & $4\pi/5$ \\
\hline
\multirow{2}{2.5em}{$DEF$} & \multirow{2}{2em}{0.55}  & 0.45 & 0 & 0.45 & 0.55 \\
&  & $-2\pi/5$ & 0 & $\pi$ & $-\pi/5$\\
\hline
\end{tabular}
\caption{Optimal states saturating the bound in Eq.~(\ref{eursurprise}). This bound can be achieved only by the triplets in $\mathcal{S}_2$, listed here in the first column. We report one state for each triplet, having set $\phi_0 = 0$.}
 \label{tab}
\end{table} 
\end{center}
We report in Table~\ref{tab} a list of optimal states saturating the bound in Eq.~(\ref{eursurprise}), one for each triplet in $\mathcal{S}_2$. We show in the appendix that these states are all related to each other by specific unitary transformations. Moreover, we prove there that triplets belonging to the same set feature the same entropic uncertainty.
\\
In the case $d=4$, there is a three-parameter family of triplets of MUBs \cite{brierley2010}, but just for a specific choice of the parameters a triplet can be extended to the unique complete set, considered in Ref.~\cite{riccardi2017}. Triplets selected from the complete set are all equivalent. Indeed, in this case
we could not find a similar mismatch in the 
URs for three MUBs: in the complete set, for $d=4$ there are five MUBs connected to five observables ($\hat{A},\hat{B},\hat{C},\hat{D},\hat{E}$), and we checked all $\binom{5}{3}=10$ combinations of triplets $\hat{X},\hat{Y},\hat{Z}$ in $\{\hat{A},\hat{B},\hat{C},\hat{D},\hat{E}\}$. They all have the lower bound $H(\hat{X})+H(\hat{Y})+H(\hat{Z})\geqslant 3$ identified in Ref.~\cite{riccardi2017}. 
It is worth mentioning that, while in the case $d=3$ the states achieving the lower bound are the same for every triplet, in $d=4$ the optimal states depend on the choice of the MUBs. We note that they share a common structure, i.e. they all have two non-null components.
\\
Interestingly, if one does not fix the parameters of the family of triplets of MUBs in $d=4$ to get a complete set, a richer scenario of inequivalent triplets depending on three parameters can be explored. Then, different URs may be obtained, with lower bounds presumably depending on the parameters. We will not explore this case here, as our analysis focuses on the complementary properties of a same set of MUBs, but we point it out as a relevant issue for further research.

By studying the sum of variances in $d=4$ and $d=5$, we find 
URs featuring the same behavior as the one found for the EURs: In $d=4$ the optimal states attain the same minimum for every choice of the bases, while in $d=5$ there are two distinct lower bounds dividing the triplets of MUBs in the same two sets as for the EURs.  
For $d=4$ all the 
URs for every choice of the three MUBs read
\begin{equation}
    \Delta X^2+\Delta Y^2+\Delta Z^2\geq 0.75 
\end{equation}
with $\hat{X}\neq \hat{Y}\neq \hat{Z} \in \{\hat{A},\hat{B},\hat{C},\hat{D},\hat{E}\}$.
\\
For $d=5$, the 
UR corresponding to the EUR in Eq.~\eqref{euralberto} reads
\begin{equation}
\Delta X^2+\Delta Y^2+\Delta Z^2\gtrsim 1.67, \label{variances1}
\end{equation}
with $XYZ\in\mathcal{S}_1$
while the one related to the triplets in Eq.~\eqref{eursurprise} is
\begin{equation}
\Delta X^2+\Delta Y^2+\Delta Z^2\gtrsim 1.37,
\label{variances2}
\end{equation}
with $XYZ\in\mathcal{S}_2$.

\begin{figure}
    \centering
    \includegraphics[width=\linewidth]{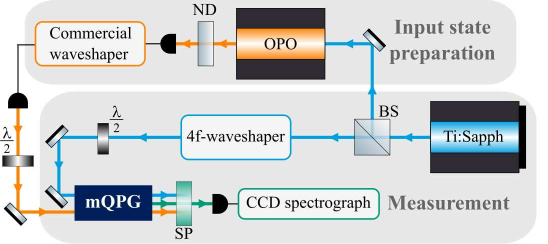}
    \caption{Schematic of the experimental setup. The signal (\SI{1545}{\nm}) and pump (\SI{860}{\nm}) pulses are generated by a Ti:Sapphire ultrafast laser with an optical parametric oscillator (OPO) at a repetition rate of \SI{80}{\MHz}. Two waveshapers generate the frequency-bin states in input from the signal pulse and the frequency-bin basis for the measurement from the pump pulse, respectively. In the mQPG waveguide, the signal modes are up-converted into a distinct output frequency based on their overlap with each pump mode. The output beam (\SI{552}{nm}) is separated from the unconverted signal and pump beams by a shortpass filter (SP) and then detected by a commercial CCD spectrograph (Andor Shamrock 500i).}
    \label{fig:setup}
\end{figure}

\section{Experimental verification}
We experimentally tested the entropic results presented in Eqs. \eqref{euralberto} and \eqref{eursurprise} by encoding information in photonic time-frequency modes \cite{brecht2015}. Namely, we consider a
Hilbert space generated by broadband frequency bins and their
superpositions encoded in coherent light pulses. In this encoding alphabet,
the five-dimensional computational basis $|a_j\>$ (associated to observable $\hat{A}$) is defined as a set of five
Gaussian-shaped frequency bins centered at different frequencies, and
the MUBs are generated by superimposing the fundamental bins with
different phases. We note that the chosen alphabet falls in the
category of pulsed temporal modes as the superposition states
overlap in both time and frequency.

We perform the projective measurements using a so-called multi-output quantum pulse gate (mQPG) \cite{serino2023}, a high-dimensional decoder for time-frequency pulsed modes based on sum-frequency generation in a dispersion-engineered waveguide. This device projects a high-dimensional input state onto all the eigenstates of a user-chosen MUB, selected via spectral shaping of the pump pulse driving the process, and yields the result of each projection in the corresponding output channel defined by a distinct output frequency.

Mathematically, we can describe the mQPG operation as a positive-operator-valued measure (POVM) $\{\pi^\gamma\}$, where each POVM element $\pi^\gamma$ describes the measurement operator of a single channel set to detect mode $\gamma$ \cite{brecht2014, ansari2017, serino2023}. Ideally $\pi^\gamma=\ketbra{\gamma}$; however, experimental imperfections lead to systematic errors in the POVMs, necessitating the more general description $\pi^\gamma=\sum_{ij}m^\gamma_{ij}\ket{a_i}\bra{a_j}$, with $\ket{a_i}$ and $\ket{a_j}$ eigenstates of the computational basis. When measuring a pure input state $\rho^\xi=\ketbra{\xi}$ from the chosen five-dimensional Hilbert space, we will obtain output $\gamma$ with probability $p^{\gamma\xi} = \mathrm{Tr}(\rho^\xi \pi^\gamma)$. 
In the measurement process, we use an mQPG with five channels, which can be programmed to perform projections onto any arbitrary MUB in the selected Hilbert space by assigning to each channel an eigenstate of that basis. 
Then, for each probed input state $\ket{\xi}$, we calculate the entropy in each measurement basis by estimating the probability $p_j$ for each measurement outcome $j$ as the normalized counts in the corresponding channel.

A schematic of the experimental setup is shown in Fig.~\ref{fig:setup}. The signal and pump pulses, centered at \SI{1545}{\nm} and \SI{860}{\nm} respectively, are generated by a combination of a Ti:Sapphire ultrafast laser with an optical parametric oscillator (OPO) at a repetition rate of \SI{80}{\MHz}. The signal pulse is shaped by a commercial waveshaper to generate the frequency-bin states in input, whereas the pump pulse is shaped by a in-house-built 4f-waveshaper to generate the frequency-bin basis for the measurement. Both beams are coupled into the mQPG waveguide, where the signal modes are up-converted into a different output channel (corresponding to a distinct output frequency) based on their overlap with each pump mode, i.e., with each eigenstate of the chosen measurement basis. The output beam, centered at around  \SI{552}{nm}, is separated from the unconverted signal and pump beams by a shortpass filter and then detected by a commercial CCD spectrograph (Andor Shamrock 500i). The number of counts detected at each output frequency indicates the number of photons measured in the corresponding mode.

\begin{figure*}[htb]
    \centering
    \includegraphics{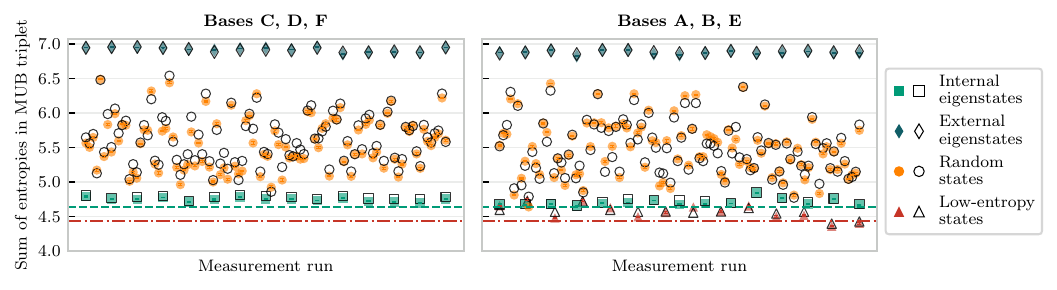}
    \caption{Sum of the entropies calculated in $d=5$ for the two MUB triplets $CDF$ (left) and $ABE$ (right) for different types of input states: eigenstates of the MUBs in the selected triplet (green squares), eigenstates of the other MUBs (blue diamonds), random states (yellow circles), and low-entropy states that violate the previous assumption of bound \eqref{euralberto} (red triangles). The filled markers show the experimental data, whereas the hollow markers describe the predicted results based on the characterized imperfect POVMs. The dashed green line and dash-dotted red line indicate the two lower bounds \eqref{euralberto} and \eqref{eursurprise}, respectively.}
    \label{fig:scatterplot}
\end{figure*}


\section{Results and discussion}
For the experimental verification of the bounds in Eqs. ~\eqref{euralberto} and~\eqref{eursurprise}, we probed different types of input states and, for each state, we calculated the entropy sum $H(\hat{X})+H(\hat{Y})+H(\hat{Z})$ from the measured output probabilities. The results for the two MUB triplets $CDF$ and $ABE$ are shown as filled markers in Figure ~\ref{fig:scatterplot}, and compared to their respective lower entropy bounds from Eqs.~\eqref{euralberto} (green dashed line) and~\eqref{eursurprise} (red dash-dotted line). For each input state, the error bars are calculated by sampling 500 sets of counts from a distribution with the measured mean and standard deviation of the original dataset and taking the 10\%-90\% spread of the corresponding entropy distribution. The error bars are not visible in most points due to their narrow extent.

The figure also shows the entropy values predicted via realistic simulations of the measurement process. These simulations are obtained by first characterizing the performance of the mQPG-based decoder through a quantum detector tomography \cite{lundeen2009, ansari2018, serino2023} to reconstruct the actual POVMs. The average measurement error (cross-talk) per basis falls between 0.5\% (for basis $\{|a_j\>\}_j$) and 1.9\% (for basis $\{|f_j\>\}_j$). From the reconstructed POVMs, we calculated the expected entropy that one would observe measuring each state with this imperfect system, obtaining the hollow markers in Fig.~\ref{fig:scatterplot}. These estimates almost perfectly match the measured values, confirming that discrepancies with the theoretical predictions are to be attributed to the imperfect detection system.

The first type of probed input states are the eigenstates of all six MUBs, labelled ``internal'' if they are eigenstates of $\hat{X}$, $\hat{Y}$ or $\hat{Z}$, and ``external'' otherwise. The entropy sum of the internal eigenstates (green squares) is always equal to the value predicted in Eq.~\eqref{euralberto} which, for the triplet $CDF$, is the minimum possible entropy sum. Contrarily, the external eigenstates (blue diamonds) always maximize the entropy sum.

Then we probed pure random input states (orange circles), generated by sampling amplitude and phase coefficients from a uniform distribution and renormalizing the amplitudes. We note that this method does not allow for truly uniform sampling of the parameter space; however, this is not relevant in the scope of this work, as we only look at random input states to verify that they fall within the predicted entropy boundaries.

Finally, we probed states that are a superposition of four states of the computational basis $|a_i\>$, e.g.\ $\sum_j\psi_je^{i\phi_j}|a_j\>$ with $\{\psi_j\}=(0.19323, 0.68019, 0, 0.68019, 0.19323)$ and $\{\phi_j\}=\left(-{3}\pi/{5}, {\pi}/{5}, 0, 0, 0\right)$ (red triangles in Fig.~\ref{fig:scatterplot}). These states were initially proven to minimize the sum of entropies in four, five and six MUBs in $d=5$ \cite{riccardi2017}. However, while testing this assumption, we observed that the sum of entropies in the $ABE$ triplet violated bound \eqref{euralberto}, previously assumed to hold for any possible choice of MUB triplet. This contradiction led to the discovery of bound in Eq.~\eqref{eursurprise} for the basis triplet $ABE$, while $CDF$ maintains the known bound, revealing the immediate experimental consequences of the underlying asymmetry between sets of inequivalent MUBs.

Interestingly, while our primary goal in this demonstration was to highlight the effect of inequivalent sets of MUBs on uncertainty relations, our findings also provide a new experimental verification of their existence that can, in principle, be extended to higher dimensions. In fact, while the existence and implications of inequivalent classes of MUBs have been theoretically understood, their experimental verification has remained elusive until very recently due to the subtlety of these effects, which could only be observed by a setup capable of performing simultaneous projections across a complete and arbitrary basis. Ref. \cite{yan2024} provided an important verification in $d=4$ using hybrid path-polarization encodings; however, their method is based on indirect estimations, and lacks straightforward scalability to higher dimensions. In contrast, our demonstration relates the existence of inequivalent sets of MUBs to the simplest and most direct type of measurements. The versatility of the mQPG operating in the time-frequency degree of freedom makes our approach naturally extensible \cite{serino2023, de24}, enabling experimental verification of similar UR bounds if they are discovered in higher-dimensional systems.

\section{Conclusions}
\label{4}
In conclusion we have shown that the URs are shaped by the inequivalence of sets of MUBs, which is a physical fundamental proof that the amount of joint information that one can have on maximally complementary observables can depend on which of them are considered, even though the overlap between the eigenstates of all maximally complementary observables are all the same, since they are MUBs. We showed that 
two distinct URs appear for triplets of MUBs in dimension
$d=5$, when the maximum joint information is gauged by minimizing the
sum of the entropies and of the variances. 


This effect has several implications which will be explored in future
work. For instance, in quantum key distribution, the entropic
uncertainty relations are known to provide a tight bound on the secret
key rate and are crucial for security proofs. In addition,
high-dimensional systems provide better key rates and larger maximum
tolerable errors. Then, complementarity-based complementarity may be
exploited to add a further level of security. Moreover, uncertainty
relations are at the basis of many quantum information procedures
(e.g. quantum metrology). Another implication refers to the
foundations of quantum mechanics, where the investigation of URs in large dimensional systems can shed light on their complementary properties.
\\
The authors acknowledge support from the EU H2020 QuantERA ERA-NET
Cofund in Quantum Technologies project QuICHE. G.C. acknowledges
support from from the PNRR MUR Project
PE0000023-NQSTI. C.M. acknowledges support from the National Research
Centre for HPC, Big Data and Quantum Computing, PNRR MUR Project
CN0000013-ICSC, and from the PRIN MUR Project
2022SW3RPY. L.M. acknowledges support from the PRIN MUR Project
2022RATBS4 and from the U.S. Department of Energy, Office of Science,
National Quantum Information Science Research Centers, Superconducting
Quantum Materials and Systems Center (SQMS) under Contract
No. DE-AC02-07CH11359.

\newpage\appendix
\begin{center} {\bf Appendix} \end{center}
\section{MUBs, explicit form} \label{s:mubs}
In the case $d=5$, six MUBs are known. The explicit expression of the
vectors for each basis is reported below as columns of the following
matrices \cite{brierley2010}:
\begin{equation} \label{mub5}
\begin{aligned}
\{|a_j\>\}_{j=1}^5 = &\begin{pmatrix}
1 & 0 & 0 & 0 & 0 \\
0 & 1 & 0 & 0 & 0 \\
0 & 0 & 1 & 0 & 0 \\
0 & 0 & 0 & 1 & 0 \\
0 & 0 & 0 & 0 & 1
\end{pmatrix} \\
\{|b_j\>\}_{j=1}^5 = \frac{1}{\sqrt{5}} &\begin{pmatrix}
1 & 1 & 1 & 1 & 1 \\
1 & \omega & \omega^2 & \omega^3 & \omega^4 \\
1 & \omega^2 & \omega^4 & \omega & \omega^3 \\
1 & \omega^3 & \omega & \omega^4 & \omega^2 \\
1 & \omega^4 & \omega^3 & \omega^2 & \omega 
\end{pmatrix} \\ 
\{|c_j\>\}_{j=1}^5= \frac{1}{\sqrt{5}} &\begin{pmatrix}
1        & 1        & 1        & 1        & 1 \\
\omega   & \omega^2 & \omega^3 & \omega^4 & 1 \\
\omega^4 & \omega   & \omega^3 & 1        & \omega^2 \\
\omega^4 & \omega^2 & 1        & \omega^3 & \omega   \\
\omega   & 1        & \omega^4 & \omega^3 & \omega^2 
                                \end{pmatrix} \\
  \{|d_j\>\}_{j=1}^5= \frac{1}{\sqrt{5}} &\begin{pmatrix}
1        & 1        & 1        & 1        & 1 \\
\omega^3 & \omega^4 & 1        & \omega   & \omega^2 \\
\omega^2 & \omega^4 & \omega   & \omega^3 & 1       \\
\omega^2 & 1        & \omega^3 & \omega   & \omega^4   \\
\omega^3 & \omega^2 & \omega   & 1        & \omega^4 
\end{pmatrix} \\
\{|e_j\>\}_{j=1}^5= \frac{1}{\sqrt{5}} &\begin{pmatrix}
1        & 1        & 1        & 1        & 1 \\
\omega^2 & \omega^3 & \omega^4 & 1        & \omega \\
\omega^3 & 1        & \omega^2 & \omega^4 & \omega \\
\omega^3 & \omega   & \omega^4 & \omega^2 & 1   \\
\omega^2 & \omega   & 1        & \omega^4 & \omega^3 
\end{pmatrix} \\
\{|f_j\>\}_{j=1}^5= \frac{1}{\sqrt{5}} &\begin{pmatrix}
1        & 1        & 1        & 1        & 1 \\
\omega^4 & 1        & \omega   & \omega^2 & \omega^3 \\
\omega   & \omega^3 & 1        & \omega^2 & \omega^4 \\
\omega   & \omega^4 & \omega^2 & 1        & \omega^3   \\
\omega^4 & \omega^3 & \omega^2 & \omega   & 1 
\end{pmatrix},
\end{aligned}
\end{equation}
where $\omega=e^{2i\pi/5}$.


In dimension four, we have five MUBs. Again, we display them as columns of the following matrices \cite{brierley2010},
\begin{equation}
\begin{aligned}
\{|a_j\>\}_{j=1}^4 = &\begin{pmatrix}
1 & 0 & 0 & 0 \\
0 & 1 & 0 & 0 \\
0 & 0 & 1 & 0 \\
0 & 0 & 0 & 1
\end{pmatrix} \\
\{|b_j\>\}_{j=1}^4 = \frac{1}{2} &\begin{pmatrix}
1 & 1  & 1   & 1  \\
1 & 1  & -1  & -1 \\
1 & -1 & -1  & 1   \\
1 & -1 & 1   & -1
\end{pmatrix} \\ 
\{|c_j\>\}_{j=1}^4= \frac{1}{2} &\begin{pmatrix}
1  & 1  & 1  & 1        \\
1  & 1  & -1 & -1  \\
-i & i  & i  & -i        \\
i  & -i & i  & -i 
\end{pmatrix} \\
\{|d_j\>\}_{j=1}^4 = \frac{1}{2} &\begin{pmatrix}
1  & 1  & 1  & 1        \\
i  & -i & i  & -i   \\
-1 & -1 & 1  & 1  \\
i  & -i & -i & i  \\
\end{pmatrix} \\
\{|e_j\>\}_{j=1}^4 = \frac{1}{2} &\begin{pmatrix}
1  & 1  & 1  & 1    \\
i  & -i & i  & -i   \\
i  & -i & -i & i \\
-1 & -1 & 1  & 1
\end{pmatrix}. 
\end{aligned}
\end{equation}

\section{Numerical minimization}\label{s:mathematica}
The numerical minimization we adopted here is very similar to the one exploited in Ref.~\cite{riccardi2017}. We used the software package \textit{Wolfram Mathematica}. We parametrized the states in dimension five as follows
\begin{equation} \label{parstate}
\begin{aligned}
|\psi\rangle =& \sin{\alpha_1}\sin{\alpha_2}\sin{\alpha_3}\sin{\alpha_4}e^{i\phi_1}|0\rangle + \\
&\cos{\alpha_1}\sin{\alpha_2}\sin{\alpha_3}\sin{\alpha_4}e^{i\phi_2}|1\rangle + \\ &\cos{\alpha_2}\sin{\alpha_3}\sin{\alpha_4}e^{i\phi_3}|2\rangle + \cos{\alpha_3}\sin{\alpha_4}e^{i\phi_4}|3\rangle + \\
&\cos{\alpha_4}e^{i\phi_5}|4\rangle,
\end{aligned}
\end{equation}
i.e. a parametrization yielding the normalization of the state intrinsically. Differently from Ref.~\cite{riccardi2017}, we checked the results of our optimizations by repeating the procedure twice: The first one, on re-parametrized states where the weights in Eq.~\eqref{parstate} were permuted with respect to the states of the basis, and the second one on each of the re-parametrized states but expanded on the eigenstates of other MUBs.
\\
For each MUB $\{|x_k\rangle\}_k$ in Eq.~\eqref{mub5} we retrieved the Born probabilities in terms of the coefficients in Eq.~\eqref{parstate} through the superpositions between the states of the MUB and the generic state $|\psi\rangle$, namely
\begin{equation}
    p(x_k) = |\langle x_k|\psi\rangle |^2.
\end{equation}
Then we considered the Shannon entropies
\begin{equation}
    H(\{|x_k\rangle\}) = -\sum_{k=1}^5p(x_k)\log_2p(x_k)
\end{equation}
and the variances
\begin{equation}
    \Delta X^2 = \langle \psi|\hat{X}^2|\psi\rangle - \langle \psi|\hat{X}|\psi\rangle^2,
\end{equation}
with $\hat{X}= \sum_k P(k)|x_k\rangle\langle x_k|$.
We then minimized numerically the sum of three entropies and the sum of three variances for all possible triplets of distinct MUBs. 
The minimization was performed by using the routine NMinimize of \textit{Wolfram Mathematica}, which finds the minimum of a function over a given set of parameters and constraints.

\section{Effects of the inequivalence of MUBs on the EURs}
Here we inspect more in detail the effect of the inequivalence of triplets of MUBs on the pertaining EURs. In particular, we show that the combination of the MUBs in the triplets from $\mathcal{S}_1$ and $\mathcal{S}_2$ identifies two related subsets of the Born probabilities inspected so far.
\\
The MUBs shown in the appendix are Hadamard matrices related to each other through the unitary $\hat{U}={\rm diag}(1,\omega,\omega^4,\omega^4,\omega)$, with $\omega = \exp{(i2\pi/5)}$, as follows
\begin{align}
    & \{|c_j\rangle\} = \hat{U}\{|b_j\rangle\} \nonumber\\
    & \{|e_j\rangle\} = \hat{U}^2\{|b_j\rangle\} \nonumber \\
    & \{|d_j\rangle\} = \hat{U}^3\{|b_j\rangle\} \nonumber \\
    & \{|f_j\rangle\} = \hat{U}^4\{|b_j\rangle\}. \nonumber
\end{align}
This structure inherently prevents from mapping any triplet in the set $\mathcal{S}_1$ of Eq.~(\ref{s1}) into a triplet in $\mathcal{S}_2$ of Eq.~(\ref{s2}) by means of a unitary matrix \cite{brierley2010}. In the following Lemma, we show how these relations impact on the Born probabilities appearing in the URs. In particular, the proof outlines how the structure of the five-dimensional MUBs implies the same EUR in Eq.~(\ref{euralberto}) (Eq.~(\ref{eursurprise})) for the triplets in the same set $\mathcal{S}_1$ ($\mathcal{S}_2$).
\begin{lemma} \label{lemma}
    Let $\mathcal{S}_1$ and $\mathcal{S}_2$ be the two inequivalent subsets of triplets of MUBs in $d=5$. 
    Be 
    \begin{equation}
        \mathcal{P}_1\equiv\{p_{X_1,\psi_0}^{(j)},p_{Y_1,\psi_0}^{(j')},p_{Z_1,\psi_0}^{(j'')}\}_{j,j',j''}
    \end{equation}
    and 
    \begin{equation}
        \mathcal{P}_2\equiv\{p_{X_2,\psi_0}^{(j)},p_{Y_2,\psi_0}^{(j')},p_{Z_2,\psi_0}^{(j'')}\}_{j,j',j''}
    \end{equation}
     the two disjoint sets of probability vectors with $X_1 Y_1 Z_1 \in \mathcal{S}_1$ and $X_2 Y_2 Z_2 \in \mathcal{S}_2$, where $p_{T,\psi_0}^{(j)}$ is the Born probability of the superposition of the $j$-th eigenstate of a basis $T \in\{A,B,C,D,E,F\}$ with a five-dimensional pure state $|\psi_0\rangle$ and $\{p_{T,\psi_0}^{(j)}\}_{j=1}^5$ is the pertaining probability vector. Consider the unitary $U={\rm diag}(1,\omega,\omega^4,\omega^4,\omega)$, with $\omega=\exp{(i2\pi/5)}$, and the Fourier matrix $\Phi \equiv \{b_j\}_j$. Then the transformations in the set $\mathcal{V}\equiv\{U^N\Phi^MU^L\}_{N,L,M}$, with $N,L \in \{0, \pm 1 \mod{5}, \pm 2 \mod{5}\}$ and $M\in\{0,\pm1\}$, are authomorphisms of $\mathcal{P}_1$ and $\mathcal{P}_2$.
\end{lemma}
\begin{proof}
The probability $p_{T,\psi_0}^{(j)}$ reads
\begin{equation}
    p_{T,\psi_0}^{(j)} = |\langle \psi_0|t_j\rangle|^2
\end{equation}
where $\{|t_j\rangle\}_j$ are the eigenstates of $T$, identifying one of the MUBs. Due to the construction of the MUBs in terms of Hadamard matrices, the probabilities $p_{T,\psi_0}^{(j)}$ can be always be expressed as
\begin{equation} \label{prob}
    p_{T,\psi_0}^{(j)} = |\langle \psi_0|U^{n_T}\Phi^{m_T}U^lP_{jk}|a_k\rangle|^2
\end{equation}
where $\{|a_j\rangle\}_j$ is the computational basis and $P_{jk}$ is the permutation exchanging the vector $|a_k\rangle$ with $|a_j\rangle$. Being $U$ a periodic matrix such that $U^5 = I$, we have $l,n_T\in \{0, \pm 1 \mod{5}, \pm 2 \mod{5}\}$, while, for the Fourier matrix, we just need $m_T = 0, \pm1$.  Permutations map the probability vectors $\{p_{T,\psi_0}^{(j)}\}_j$ in themselves and are therefore automorphisms for $\mathcal{P}_1$ and $\mathcal{P}_2$. Similarly, the action of $\hat{U}^l$ on the states of the computational basis reduces to the multiplication for a global phase factor, thus leaving $p_{T,\psi_0}^{(j)}$ unchanged. Conversely, the exponents $m_T$ and $n_T$ determines the basis $T$. If $m_T=0$, the transformation $U^{n_T}\Phi^{m_T}$, again, rescales $|a_j\rangle$ by an irrelevant phase factor, while, for $m_T=\pm 1$, maps it to the a state of another MUB, identified by $n_T$. Now, we evaluate the probabilities in Eq.~(\ref{prob}) on a state $|\psi\rangle = V|\psi_0\rangle$, with unitary $V$. By definition of $\mathcal{P}_1$ and $\mathcal{P}_2$, if $V^{\dagger}$ is an automorphism of the set of MUBs, namely $V^{\dagger}:T\rightarrow \overline{T}$, with $T$ and $\overline{T}$ mutually unbiased, as
\begin{equation}
    V^{\dagger}U^{n_T}\Phi^{m_T} = {U^{n_{\overline{T}}}}{\Phi^{m_{\overline{T}}}},
\end{equation}
then it is an automorphism of $\mathcal{P}_1\cup \mathcal{P}_2$, and
\begin{equation}
    V^{\dagger} = {U^{n_{\overline{T}}}}{\Phi^{m_{\overline{T}}-m_T}}{U^{n_T}}\equiv U^N\Phi^MU^L \in \mathcal{V}.
\end{equation}
Now we have to show that the transformations in $\mathcal{V}$ identify the two disjoint sets $\mathcal{P}_1$ and $\mathcal{P}_2$. The application of the same unitary $V$ to the state $|\psi_0\rangle$ relates the triplets of probability vectors $\{p_{X,\psi}^{(j)},p_{Y,\psi}^{(j')},p_{Z,\psi}^{(j'')}\}$ and $\{p_{\overline{X},\psi_0}^{(j)},p_{\overline{Y},\psi_0}^{(j')},p_{\overline{Z},\psi_0}^{(j'')}\}$ through the equations
\begin{align}
    &U^N\Phi^MU^{L+n_X}\Phi^{m_X}=U^{n_{\overline{X}}}\Phi^{m_{\overline{X}}} \label{eq1} \\
    &U^N\Phi^MU^{L+n_Y}\Phi^{m_Y}=U^{n_{\overline{Y}}}\Phi^{m_{\overline{Y}}} \label{eq2} \\
   &U^N\Phi^MU^{L+n_Z}\Phi^{m_Z}=U^{n_{\overline{Z}}}\Phi^{m_{\overline{Z}}} \label{eq3} \\
\end{align}
Without loss of generality, we can restrict to the cases where $M=0, L\neq 0$ and $M\neq0, L=0$, being the remaining cases retrieved by shifting $n_X$, $n_Y$ and $n_Z$ by $L$.  For the sake of simplicity, we will set the triplet $XYZ$ in the so called standard form, where one of the MUBs is the computational basis, say $ X = \{|a_j\rangle\}_j$, implying $m_X = 0$. 
The proof with $m_X=\pm 1$ follows the same line of argument and leads to the same conclusions. \\
In the case $M=0$, note that we can also set $N=0$. Then, Eqs.~(\ref{eq1}),~(\ref{eq2}) and~(\ref{eq3}) fix $m_{\overline{X}} = m_X =0$, $m_{\overline{Y}} = m_Y \neq 0$ and $m_{\overline{Z}} = m_Z \neq 0$, respectively. Note that the last two must be both non-null, or we would have repetitions of the computational basis in the same triplet. Finally, Eqs.~(\ref{eq2}) and~(\ref{eq3}) set $\Delta n\equiv n_Z - n_Y = n_{\overline{Z}} - n_{\overline{Y}}$, implying that the application of $V$ fixes a relation between $\{Y, Z\}$ and $\{\overline{Y}, \overline{Z}\}$. From the periodicity of $U$, again $\Delta n \in \{0, \pm 1 \mod{5}, \pm 2 \mod{5}\}$. We must have $\Delta n \neq 0$ to avoid $Y$ and $Z$ being the same basis. Moreover, the order in the difference $\Delta n$ is irrelevant, and so is its sign. These considerations leave us with two distinct possibilities, either $\Delta n = \pm 1 \mod{5}$ or $\Delta n = \pm 2 \mod{5}$. But these identify two disjoint sets of triplets, the one in $\mathcal{S}_1$, featuring $\Delta n = \pm 1 \mod{5}$, and the one in $\mathcal{S}_2$, with $\Delta n = \pm 2 \mod{5}$. If all the bases in the triplet are assumed to be different from $\{a_j\}_j$, then it is simple to see that one ends with a similar set of conditions for $n_X$, $n_Y$ and $n_Z$ separating the two sets of triplets as before. Hence, $U^L$ maps the elements of $\mathcal{P}_1$ in elements of $\mathcal{P}_1$ and the elements of $\mathcal{P}_2$ in elements of $\mathcal{P}_2$. 
\\
In the case $M\neq 0, L = 0$, we need the following identities,
\begin{align}
    &\Phi^{\dagger}U\Phi = U\Phi P_{jk}U \label{sym1} \\
    &\Phi^{\dagger}U^2\Phi = U^3\Phi U^2P'_{jk}. \label{sym2}
\end{align}
Again, we will neglect the contributions of permutations and global phase shifts. From Eq.~(\ref{eq1}), we have $m_{\overline{X}}=M$ and $n_{\overline{X}}=N$. By exploiting Eqs.~(\ref{sym1}) and~(\ref{sym2}), we find that the left members of Eqs.~(\ref{eq2}) and~(\ref{eq3}) yield
\begin{align}
    &U^{n_{\overline{X}}}\Phi^{m_{\overline{X}}}U^{n_k}\Phi^{m_k} = \delta_{n_k,0}(1-\delta_{n_j,0})U^{n_{\overline{X}}} + \delta_{n_k,\pm1} \nonumber \\
    &(1-\delta_{n_j,\pm1})U^{n_{\overline{X}}+n_k}\Phi^{m_k}+\delta_{n_k,2}(1-\delta_{n_j,2})\delta_{m_{\overline{X}},-1} \nonumber \\
     &U^{n_{\overline{X}}+n_k+1}\Phi^{m_k}+\delta_{n_k,-2}(1-\delta_{n_j,-2})\delta_{m_{\overline{X}},1}U^{n_{\overline{X}}+n_k-1} \nonumber \\
    &\Phi^{m_k}
\end{align}
where $\delta_{a,b}$ is the Kronecker delta, $k\neq j$ and $k,j \in \{Y,Z\}$. By requiring the equality with the right members of Eqs.~(\ref{eq2}) and~(\ref{eq3}), we find that the deltas identify the following cases:
\begin{itemize}
    \item 
        $n_{\overline{Y}}=n_{\overline{X}} \pm 1 \mod{5}\nonumber \\
        n_{\overline{Z}}=n_{\overline{X}} \mp 1 \mod{5}\nonumber$ \\
    implying $\Delta n = n_Z - n_Y = n_{\overline{Z}} - n_{\overline{Y}} = \pm 2 \mod{5}$: the transformations are automorphisms of $\mathcal{P}_2$;
    \item $n_{\overline{Y}}=n_{\overline{X}} \pm 1 \mod{5}\nonumber \\
        n_{\overline{Z}}=n_{\overline{X}} -2 \mod{5} \nonumber$ \\
        implying $\Delta n = n_{\overline{Z}} - n_{\overline{Y}} - 1$ and we have that either $\Delta n = 1 \mod{5}$ or $\Delta n = -2 \mod{5}$; in the first case, the transformations are automorphisms of $\mathcal{P}_1$, while, in the second one, of $\mathcal{P}_2$; 
        \item $n_{\overline{Y}}=n_{\overline{X}} \pm 1 \mod{5}\nonumber \\
        n_{\overline{Z}}=n_{\overline{X}} +2 \mod{5} \nonumber$ \\
        implying $\Delta n = n_{\overline{Z}} - n_{\overline{Y}} + 1$ and we have that either $\Delta n = -1 \mod{5}$ or $\Delta n = 2 \mod{5}$; in the first case, the transformations are automorphisms of $\mathcal{P}_1$, while, in the second one, of $\mathcal{P}_2$;
        \item $n_{\overline{Y}}=n_{\overline{X}} -2 \mod{5} \nonumber \\
        n_{\overline{Z}}=n_{\overline{X}} +2 \mod{5} \nonumber$ \\
        implying $\Delta n = n_{\overline{Z}} - n_{\overline{Y}} + 2 = 1 \mod{5}$: the transformations are automorphisms of $\mathcal{P}_1$.
\end{itemize}
Then, the automorphisms of the whole set of triplets of probability vectors $\mathcal{P}_1 \cup \mathcal{P}_2$ are automorphisms of the disjoint sets $\mathcal{P}_1$ and $\mathcal{P}_2$.
\end{proof}
Note that, while it has been proved \cite{brierley2010} that the elements of $\mathcal{S}_1$ cannot be mapped into elements of $\mathcal{S}_2$ via unitary transformations, Lemma~\ref{lemma} does not imply that there are no unitaries mapping elements of $\mathcal{P}_1$ into elements of $\mathcal{P}_2$. This is a relevant distinction between the sets $\mathcal{S}_1$, $\mathcal{S}_2$ and the sets $\mathcal{P}_1$, $\mathcal{P}_2$. Lemma~\ref{lemma} establish that $\mathcal{P}_1$ and $\mathcal{P}_2$ are distinct equivalence classes, which does not rule out that their elements can be unitarily related. For instance, this is trivially the case if $|\psi_0\rangle$ and $|\psi\rangle=V|\psi_0\rangle$ are states of MUBs: given two inequivalent triplets of MUBs, one can always pick two states from different MUBs providing the same triplets of probability vectors. 
\\
Since the Shannon entropies in the EURs here addressed depend on the Born probabilities only, Lemma~\ref{lemma} is a sufficient condition for the EUR in Eq.~(\ref{euralberto}) (Eq.~(\ref{eursurprise})) to hold for all the triplets in the set $\mathcal{S}_1$ ($\mathcal{S}_2$). On the other hand, the existence of distinct lower bounds is a necessary, but not sufficient, condition for the inequivalence of $\mathcal{P}_1$ and $\mathcal{P}_2$, since it implies that there is no unitary transformation that can map the optimal states saturating the EUR for the triplets in $\mathcal{S}_2$ into equivalent optimal states for the triplets in $\mathcal{S}_1$.

\bibliographystyle{apsrev4-1}
\bibliography{bib}

\begin{thebibliography}{32}%
\makeatletter
\providecommand \@ifxundefined [1]{%
 \@ifx{#1\undefined}
}%
\providecommand \@ifnum [1]{%
 \ifnum #1\expandafter \@firstoftwo
 \else \expandafter \@secondoftwo
 \fi
}%
\providecommand \@ifx [1]{%
 \ifx #1\expandafter \@firstoftwo
 \else \expandafter \@secondoftwo
 \fi
}%
\providecommand \natexlab [1]{#1}%
\providecommand \enquote  [1]{``#1''}%
\providecommand \bibnamefont  [1]{#1}%
\providecommand \bibfnamefont [1]{#1}%
\providecommand \citenamefont [1]{#1}%
\providecommand \href@noop [0]{\@secondoftwo}%
\providecommand \href [0]{\begingroup \@sanitize@url \@href}%
\providecommand \@href[1]{\@@startlink{#1}\@@href}%
\providecommand \@@href[1]{\endgroup#1\@@endlink}%
\providecommand \@sanitize@url [0]{\catcode `\\12\catcode `\$12\catcode `\&12\catcode `\#12\catcode `\^12\catcode `\_12\catcode `\%12\relax}%
\providecommand \@@startlink[1]{}%
\providecommand \@@endlink[0]{}%
\providecommand \url  [0]{\begingroup\@sanitize@url \@url }%
\providecommand \@url [1]{\endgroup\@href {#1}{\urlprefix }}%
\providecommand \urlprefix  [0]{URL }%
\providecommand \Eprint [0]{\href }%
\providecommand \doibase [0]{http://dx.doi.org/}%
\providecommand \selectlanguage [0]{\@gobble}%
\providecommand \bibinfo  [0]{\@secondoftwo}%
\providecommand \bibfield  [0]{\@secondoftwo}%
\providecommand \translation [1]{[#1]}%
\providecommand \BibitemOpen [0]{}%
\providecommand \bibitemStop [0]{}%
\providecommand \bibitemNoStop [0]{.\EOS\space}%
\providecommand \EOS [0]{\spacefactor3000\relax}%
\providecommand \BibitemShut  [1]{\csname bibitem#1\endcsname}%
\let\auto@bib@innerbib\@empty
\bibitem [{\citenamefont {Bohr}(1928{\natexlab{a}})}]{bohr1928}%
  \BibitemOpen
  \bibfield  {author} {\bibinfo {author} {\bibfnamefont {N.}~\bibnamefont {Bohr}},\ }\href@noop {} {\bibfield  {journal} {\bibinfo  {journal} {Naturwissenschaften}\ }\textbf {\bibinfo {volume} {16}},\ \bibinfo {pages} {245} (\bibinfo {year} {1928}{\natexlab{a}})}\BibitemShut {NoStop}%
\bibitem [{\citenamefont {Bohr}(1928{\natexlab{b}})}]{bohrbohr1928}%
  \BibitemOpen
  \bibfield  {author} {\bibinfo {author} {\bibfnamefont {N.}~\bibnamefont {Bohr}},\ }\href@noop {} {\bibfield  {journal} {\bibinfo  {journal} {Nature}\ }\textbf {\bibinfo {volume} {121}},\ \bibinfo {pages} {580} (\bibinfo {year} {1928}{\natexlab{b}})}\BibitemShut {NoStop}%
\bibitem [{\citenamefont {Wootters}\ and\ \citenamefont {Zurek}(1979)}]{wootters1979}%
  \BibitemOpen
  \bibfield  {author} {\bibinfo {author} {\bibfnamefont {W.~K.}\ \bibnamefont {Wootters}}\ and\ \bibinfo {author} {\bibfnamefont {W.~H.}\ \bibnamefont {Zurek}},\ }\href@noop {} {\bibfield  {journal} {\bibinfo  {journal} {Phys. Rev. D}\ }\textbf {\bibinfo {volume} {19}},\ \bibinfo {pages} {473} (\bibinfo {year} {1979})}\BibitemShut {NoStop}%
\bibitem [{\citenamefont {Greenberger}\ and\ \citenamefont {Yasin}(1988)}]{greenberger1988}%
  \BibitemOpen
  \bibfield  {author} {\bibinfo {author} {\bibfnamefont {D.~M.}\ \bibnamefont {Greenberger}}\ and\ \bibinfo {author} {\bibfnamefont {A.}~\bibnamefont {Yasin}},\ }\href@noop {} {\bibfield  {journal} {\bibinfo  {journal} {Phys. Lett. A}\ }\textbf {\bibinfo {volume} {128}},\ \bibinfo {pages} {391} (\bibinfo {year} {1988})}\BibitemShut {NoStop}%
\bibitem [{\citenamefont {Englert}(1996)}]{englert1996}%
  \BibitemOpen
  \bibfield  {author} {\bibinfo {author} {\bibfnamefont {B.-G.}\ \bibnamefont {Englert}},\ }\href@noop {} {\bibfield  {journal} {\bibinfo  {journal} {Phys. Rev. Lett.}\ }\textbf {\bibinfo {volume} {77}},\ \bibinfo {pages} {2154} (\bibinfo {year} {1996})}\BibitemShut {NoStop}%
\bibitem [{\citenamefont {Tadej}\ and\ \citenamefont {Zyczkowski}(2006)}]{tadej2006}%
  \BibitemOpen
  \bibfield  {author} {\bibinfo {author} {\bibfnamefont {W.}~\bibnamefont {Tadej}}\ and\ \bibinfo {author} {\bibfnamefont {K.}~\bibnamefont {Zyczkowski}},\ }\href@noop {} {\bibfield  {journal} {\bibinfo  {journal} {Open Syst. Inf. Dyn.}\ }\textbf {\bibinfo {volume} {13}},\ \bibinfo {pages} {133} (\bibinfo {year} {2006})}\BibitemShut {NoStop}%
\bibitem [{\citenamefont {Durt}\ \emph {et~al.}(2010)\citenamefont {Durt}, \citenamefont {Englert}, \citenamefont {Bengtsson},\ and\ \citenamefont {\.Zyczkowski}}]{durt2010}%
  \BibitemOpen
  \bibfield  {author} {\bibinfo {author} {\bibfnamefont {T.}~\bibnamefont {Durt}}, \bibinfo {author} {\bibfnamefont {B.-G.}\ \bibnamefont {Englert}}, \bibinfo {author} {\bibfnamefont {I.}~\bibnamefont {Bengtsson}}, \ and\ \bibinfo {author} {\bibnamefont {\.Zyczkowski}},\ }\href@noop {} {\bibfield  {journal} {\bibinfo  {journal} {Int. J. Quantum Inf.}\ }\textbf {\bibinfo {volume} {8}},\ \bibinfo {pages} {535} (\bibinfo {year} {2010})}\BibitemShut {NoStop}%
\bibitem [{\citenamefont {Brierley}\ \emph {et~al.}(2010)\citenamefont {Brierley}, \citenamefont {Weigert},\ and\ \citenamefont {Bengtsson}}]{brierley2010}%
  \BibitemOpen
  \bibfield  {author} {\bibinfo {author} {\bibfnamefont {S.}~\bibnamefont {Brierley}}, \bibinfo {author} {\bibfnamefont {S.}~\bibnamefont {Weigert}}, \ and\ \bibinfo {author} {\bibfnamefont {I.}~\bibnamefont {Bengtsson}},\ }\href@noop {} {\bibfield  {journal} {\bibinfo  {journal} {Quantum Inf. Comput.}\ }\textbf {\bibinfo {volume} {10}},\ \bibinfo {pages} {0803} (\bibinfo {year} {2010})}\BibitemShut {NoStop}%
\bibitem [{\citenamefont {Kantor}(2012)}]{kantor2012}%
  \BibitemOpen
  \bibfield  {author} {\bibinfo {author} {\bibfnamefont {W.~M.}\ \bibnamefont {Kantor}},\ }\href@noop {} {\bibfield  {journal} {\bibinfo  {journal} {J. Math. Phys.}\ }\textbf {\bibinfo {volume} {53}},\ \bibinfo {pages} {032204} (\bibinfo {year} {2012})}\BibitemShut {NoStop}%
\bibitem [{\citenamefont {Sehrawat}\ and\ \citenamefont {Klimov}(2014)}]{sehrawat2014}%
  \BibitemOpen
  \bibfield  {author} {\bibinfo {author} {\bibfnamefont {A.}~\bibnamefont {Sehrawat}}\ and\ \bibinfo {author} {\bibfnamefont {A.~B.}\ \bibnamefont {Klimov}},\ }\href@noop {} {\bibfield  {journal} {\bibinfo  {journal} {Phys. Rev. A}\ }\textbf {\bibinfo {volume} {90}},\ \bibinfo {pages} {062308} (\bibinfo {year} {2014})}\BibitemShut {NoStop}%
\bibitem [{\citenamefont {Designolle}\ \emph {et~al.}(2019)\citenamefont {Designolle}, \citenamefont {Skrzypczyk}, \citenamefont {Fr\"{o}wis},\ and\ \citenamefont {Brunner}}]{designolle2019}%
  \BibitemOpen
  \bibfield  {author} {\bibinfo {author} {\bibfnamefont {S.}~\bibnamefont {Designolle}}, \bibinfo {author} {\bibfnamefont {P.}~\bibnamefont {Skrzypczyk}}, \bibinfo {author} {\bibfnamefont {F.}~\bibnamefont {Fr\"{o}wis}}, \ and\ \bibinfo {author} {\bibfnamefont {N.}~\bibnamefont {Brunner}},\ }\href@noop {} {\bibfield  {journal} {\bibinfo  {journal} {Phys. Rev. Lett.}\ }\textbf {\bibinfo {volume} {122}},\ \bibinfo {pages} {050402} (\bibinfo {year} {2019})}\BibitemShut {NoStop}%
\bibitem [{\citenamefont {Aguilar}\ \emph {et~al.}(2018)\citenamefont {Aguilar}, \citenamefont {Borka{\l}a}, \citenamefont {Mironowicz},\ and\ \citenamefont {Paw{\l}owski}}]{aguilar2018}%
  \BibitemOpen
  \bibfield  {author} {\bibinfo {author} {\bibfnamefont {E.~A.}\ \bibnamefont {Aguilar}}, \bibinfo {author} {\bibfnamefont {J.~J.}\ \bibnamefont {Borka{\l}a}}, \bibinfo {author} {\bibfnamefont {P.}~\bibnamefont {Mironowicz}}, \ and\ \bibinfo {author} {\bibfnamefont {M.}~\bibnamefont {Paw{\l}owski}},\ }\href@noop {} {\bibfield  {journal} {\bibinfo  {journal} {Phys. Rev. Lett.}\ }\textbf {\bibinfo {volume} {121}},\ \bibinfo {pages} {050501} (\bibinfo {year} {2018})}\BibitemShut {NoStop}%
\bibitem [{\citenamefont {Hiesmayr}\ \emph {et~al.}(2021)\citenamefont {Hiesmayr}, \citenamefont {McNulty}, \citenamefont {Baek}, \citenamefont {Singha~Roy}, \citenamefont {Bae},\ and\ \citenamefont {Chru\'sci\'nski}}]{hiesmayr2021}%
  \BibitemOpen
  \bibfield  {author} {\bibinfo {author} {\bibfnamefont {B.~C.}\ \bibnamefont {Hiesmayr}}, \bibinfo {author} {\bibfnamefont {D.}~\bibnamefont {McNulty}}, \bibinfo {author} {\bibfnamefont {S.}~\bibnamefont {Baek}}, \bibinfo {author} {\bibfnamefont {S.}~\bibnamefont {Singha~Roy}}, \bibinfo {author} {\bibfnamefont {J.}~\bibnamefont {Bae}}, \ and\ \bibinfo {author} {\bibfnamefont {D.}~\bibnamefont {Chru\'sci\'nski}},\ }\href@noop {} {\bibfield  {journal} {\bibinfo  {journal} {New J. Phys.}\ }\textbf {\bibinfo {volume} {23}},\ \bibinfo {pages} {093018} (\bibinfo {year} {2021})}\BibitemShut {NoStop}%
\bibitem [{Note1()}]{Note1}%
  \BibitemOpen
  \bibinfo {note} {We also considered the Renyi entropies of order 2 and 1/2, but the numerics gave inconclusive results}\BibitemShut {NoStop}%
\bibitem [{\citenamefont {Deutsch}(1983)}]{deutsch1983}%
  \BibitemOpen
  \bibfield  {author} {\bibinfo {author} {\bibfnamefont {D.}~\bibnamefont {Deutsch}},\ }\href@noop {} {\bibfield  {journal} {\bibinfo  {journal} {Phys. Rev. Lett.}\ }\textbf {\bibinfo {volume} {50}},\ \bibinfo {pages} {631} (\bibinfo {year} {1983})}\BibitemShut {NoStop}%
\bibitem [{\citenamefont {Maassen}\ and\ \citenamefont {Uffink}(1988)}]{maassen1988}%
  \BibitemOpen
  \bibfield  {author} {\bibinfo {author} {\bibfnamefont {H.}~\bibnamefont {Maassen}}\ and\ \bibinfo {author} {\bibfnamefont {J.~B.~M.}\ \bibnamefont {Uffink}},\ }\href@noop {} {\bibfield  {journal} {\bibinfo  {journal} {Phys. Rev. Lett.}\ }\textbf {\bibinfo {volume} {60}},\ \bibinfo {pages} {1103} (\bibinfo {year} {1988})}\BibitemShut {NoStop}%
\bibitem [{\citenamefont {Riccardi}\ \emph {et~al.}(2017)\citenamefont {Riccardi}, \citenamefont {Macchiavello},\ and\ \citenamefont {Maccone}}]{riccardi2017}%
  \BibitemOpen
  \bibfield  {author} {\bibinfo {author} {\bibfnamefont {A.}~\bibnamefont {Riccardi}}, \bibinfo {author} {\bibfnamefont {C.}~\bibnamefont {Macchiavello}}, \ and\ \bibinfo {author} {\bibfnamefont {L.}~\bibnamefont {Maccone}},\ }\href@noop {} {\bibfield  {journal} {\bibinfo  {journal} {Phys. Rev. A}\ }\textbf {\bibinfo {volume} {95}},\ \bibinfo {pages} {032109} (\bibinfo {year} {2017})}\BibitemShut {NoStop}%
\bibitem [{\citenamefont {Trifonov}\ and\ \citenamefont {Donev}(1998)}]{trifonov1998}%
  \BibitemOpen
  \bibfield  {author} {\bibinfo {author} {\bibfnamefont {D.~A.}\ \bibnamefont {Trifonov}}\ and\ \bibinfo {author} {\bibfnamefont {S.~G.}\ \bibnamefont {Donev}},\ }\href@noop {} {\bibfield  {journal} {\bibinfo  {journal} {J. Phys. A: Math. Gen.}\ }\textbf {\bibinfo {volume} {31}},\ \bibinfo {pages} {8041} (\bibinfo {year} {1998})}\BibitemShut {NoStop}%
\bibitem [{Note2()}]{Note2}%
  \BibitemOpen
  \bibinfo {note} {Instead, the product of variances, as in the traditional Heisenberg-Robertson uncertainty relations \cite {robertson1929}, is problematic because it has no nonzero lower bound when minimizing over states: the product is null already if only one of the two variances is nonzero.}\BibitemShut {Stop}%
\bibitem [{\citenamefont {Trifonov}(2000)}]{trifonov2000}%
  \BibitemOpen
  \bibfield  {author} {\bibinfo {author} {\bibfnamefont {D.~A.}\ \bibnamefont {Trifonov}},\ }\href@noop {} {\bibfield  {journal} {\bibinfo  {journal} {J. Opt. Soc. Am. A}\ }\textbf {\bibinfo {volume} {17}},\ \bibinfo {pages} {2486} (\bibinfo {year} {2000})}\BibitemShut {NoStop}%
\bibitem [{\citenamefont {Trifonov}(2002)}]{trifonov2002}%
  \BibitemOpen
  \bibfield  {author} {\bibinfo {author} {\bibfnamefont {D.~A.}\ \bibnamefont {Trifonov}},\ }\href@noop {} {\bibfield  {journal} {\bibinfo  {journal} {Eur. Phys. J. B}\ }\textbf {\bibinfo {volume} {29}},\ \bibinfo {pages} {349} (\bibinfo {year} {2002})}\BibitemShut {NoStop}%
\bibitem [{\citenamefont {Zyczkowski}\ and\ \citenamefont {Ku\'{s}}(1994)}]{zyczkowski1994}%
  \BibitemOpen
  \bibfield  {author} {\bibinfo {author} {\bibfnamefont {K.}~\bibnamefont {Zyczkowski}}\ and\ \bibinfo {author} {\bibfnamefont {M.}~\bibnamefont {Ku\'{s}}},\ }\href@noop {} {\bibfield  {journal} {\bibinfo  {journal} {J. Phys. A: Math. Gen.}\ }\textbf {\bibinfo {volume} {27}},\ \bibinfo {pages} {4235} (\bibinfo {year} {1994})}\BibitemShut {NoStop}%
\bibitem [{\citenamefont {Po\'{z}niak}\ \emph {et~al.}(1998)\citenamefont {Po\'{z}niak}, \citenamefont {Zyczkowski},\ and\ \citenamefont {Ku\'{s}}}]{zyczkowski1998}%
  \BibitemOpen
  \bibfield  {author} {\bibinfo {author} {\bibfnamefont {M.}~\bibnamefont {Po\'{z}niak}}, \bibinfo {author} {\bibfnamefont {K.}~\bibnamefont {Zyczkowski}}, \ and\ \bibinfo {author} {\bibfnamefont {M.}~\bibnamefont {Ku\'{s}}},\ }\href@noop {} {\bibfield  {journal} {\bibinfo  {journal} {J. Phys. A: Math. Gen.}\ }\textbf {\bibinfo {volume} {31}},\ \bibinfo {pages} {1059} (\bibinfo {year} {1998})}\BibitemShut {NoStop}%
\bibitem [{\citenamefont {Brecht}\ \emph {et~al.}(2015)\citenamefont {Brecht}, \citenamefont {Reddy}, \citenamefont {Silberhorn},\ and\ \citenamefont {Raymer}}]{brecht2015}%
  \BibitemOpen
  \bibfield  {author} {\bibinfo {author} {\bibfnamefont {B.}~\bibnamefont {Brecht}}, \bibinfo {author} {\bibfnamefont {D.~V.}\ \bibnamefont {Reddy}}, \bibinfo {author} {\bibfnamefont {C.}~\bibnamefont {Silberhorn}}, \ and\ \bibinfo {author} {\bibfnamefont {M.~G.}\ \bibnamefont {Raymer}},\ }\href {\doibase 10.1103/PhysRevX.5.041017} {\bibfield  {journal} {\bibinfo  {journal} {Phys. Rev. X}\ }\textbf {\bibinfo {volume} {5}},\ \bibinfo {pages} {041017} (\bibinfo {year} {2015})}\BibitemShut {NoStop}%
\bibitem [{\citenamefont {Serino}\ \emph {et~al.}(2023)\citenamefont {Serino}, \citenamefont {Gil-Lopez}, \citenamefont {Stefszky}, \citenamefont {Ricken}, \citenamefont {Eigner}, \citenamefont {Brecht},\ and\ \citenamefont {Silberhorn}}]{serino2023}%
  \BibitemOpen
  \bibfield  {author} {\bibinfo {author} {\bibfnamefont {L.}~\bibnamefont {Serino}}, \bibinfo {author} {\bibfnamefont {J.}~\bibnamefont {Gil-Lopez}}, \bibinfo {author} {\bibfnamefont {M.}~\bibnamefont {Stefszky}}, \bibinfo {author} {\bibfnamefont {R.}~\bibnamefont {Ricken}}, \bibinfo {author} {\bibfnamefont {C.}~\bibnamefont {Eigner}}, \bibinfo {author} {\bibfnamefont {B.}~\bibnamefont {Brecht}}, \ and\ \bibinfo {author} {\bibfnamefont {C.}~\bibnamefont {Silberhorn}},\ }\href {\doibase 10.1103/PRXQuantum.4.020306} {\bibfield  {journal} {\bibinfo  {journal} {PRX Quantum}\ }\textbf {\bibinfo {volume} {4}},\ \bibinfo {pages} {020306} (\bibinfo {year} {2023})}\BibitemShut {NoStop}%
\bibitem [{\citenamefont {Brecht}\ \emph {et~al.}(2014)\citenamefont {Brecht}, \citenamefont {Eckstein}, \citenamefont {Ricken}, \citenamefont {Quiring}, \citenamefont {Suche}, \citenamefont {Sansoni},\ and\ \citenamefont {Silberhorn}}]{brecht2014}%
  \BibitemOpen
  \bibfield  {author} {\bibinfo {author} {\bibfnamefont {B.}~\bibnamefont {Brecht}}, \bibinfo {author} {\bibfnamefont {A.}~\bibnamefont {Eckstein}}, \bibinfo {author} {\bibfnamefont {R.}~\bibnamefont {Ricken}}, \bibinfo {author} {\bibfnamefont {V.}~\bibnamefont {Quiring}}, \bibinfo {author} {\bibfnamefont {H.}~\bibnamefont {Suche}}, \bibinfo {author} {\bibfnamefont {L.}~\bibnamefont {Sansoni}}, \ and\ \bibinfo {author} {\bibfnamefont {C.}~\bibnamefont {Silberhorn}},\ }\href {\doibase 10.1103/PhysRevA.90.030302} {\bibfield  {journal} {\bibinfo  {journal} {Phys. Rev. A}\ }\textbf {\bibinfo {volume} {90}},\ \bibinfo {pages} {030302(R)} (\bibinfo {year} {2014})}\BibitemShut {NoStop}%
\bibitem [{\citenamefont {Ansari}\ \emph {et~al.}(2017)\citenamefont {Ansari}, \citenamefont {Harder}, \citenamefont {Allgaier}, \citenamefont {Brecht},\ and\ \citenamefont {Silberhorn}}]{ansari2017}%
  \BibitemOpen
  \bibfield  {author} {\bibinfo {author} {\bibfnamefont {V.}~\bibnamefont {Ansari}}, \bibinfo {author} {\bibfnamefont {G.}~\bibnamefont {Harder}}, \bibinfo {author} {\bibfnamefont {M.}~\bibnamefont {Allgaier}}, \bibinfo {author} {\bibfnamefont {B.}~\bibnamefont {Brecht}}, \ and\ \bibinfo {author} {\bibfnamefont {C.}~\bibnamefont {Silberhorn}},\ }\href {\doibase 10.1103/PhysRevA.96.063817} {\bibfield  {journal} {\bibinfo  {journal} {Phys. Rev. A}\ }\textbf {\bibinfo {volume} {96}},\ \bibinfo {pages} {063817} (\bibinfo {year} {2017})}\BibitemShut {NoStop}%
\bibitem [{\citenamefont {Lundeen}\ \emph {et~al.}(2009)\citenamefont {Lundeen}, \citenamefont {Feito}, \citenamefont {Coldenstrodt-Ronge}, \citenamefont {Pregnell}, \citenamefont {Silberhorn}, \citenamefont {Ralph}, \citenamefont {Eisert}, \citenamefont {Plenio},\ and\ \citenamefont {Walmsley}}]{lundeen2009}%
  \BibitemOpen
  \bibfield  {author} {\bibinfo {author} {\bibfnamefont {J.~S.}\ \bibnamefont {Lundeen}}, \bibinfo {author} {\bibfnamefont {A.}~\bibnamefont {Feito}}, \bibinfo {author} {\bibfnamefont {H.}~\bibnamefont {Coldenstrodt-Ronge}}, \bibinfo {author} {\bibfnamefont {K.~L.}\ \bibnamefont {Pregnell}}, \bibinfo {author} {\bibfnamefont {C.}~\bibnamefont {Silberhorn}}, \bibinfo {author} {\bibfnamefont {T.~C.}\ \bibnamefont {Ralph}}, \bibinfo {author} {\bibfnamefont {J.}~\bibnamefont {Eisert}}, \bibinfo {author} {\bibfnamefont {M.~B.}\ \bibnamefont {Plenio}}, \ and\ \bibinfo {author} {\bibfnamefont {I.~A.}\ \bibnamefont {Walmsley}},\ }\href {\doibase 10.1038/nphys1133} {\bibfield  {journal} {\bibinfo  {journal} {Nature Physics}\ }\textbf {\bibinfo {volume} {5}},\ \bibinfo {pages} {27} (\bibinfo {year} {2009})}\BibitemShut {NoStop}%
\bibitem [{\citenamefont {Ansari}\ \emph {et~al.}(2018)\citenamefont {Ansari}, \citenamefont {Donohue}, \citenamefont {Allgaier}, \citenamefont {Sansoni}, \citenamefont {Brecht}, \citenamefont {Roslund}, \citenamefont {Treps}, \citenamefont {Harder},\ and\ \citenamefont {Silberhorn}}]{ansari2018}%
  \BibitemOpen
  \bibfield  {author} {\bibinfo {author} {\bibfnamefont {V.}~\bibnamefont {Ansari}}, \bibinfo {author} {\bibfnamefont {J.~M.}\ \bibnamefont {Donohue}}, \bibinfo {author} {\bibfnamefont {M.}~\bibnamefont {Allgaier}}, \bibinfo {author} {\bibfnamefont {L.}~\bibnamefont {Sansoni}}, \bibinfo {author} {\bibfnamefont {B.}~\bibnamefont {Brecht}}, \bibinfo {author} {\bibfnamefont {J.}~\bibnamefont {Roslund}}, \bibinfo {author} {\bibfnamefont {N.}~\bibnamefont {Treps}}, \bibinfo {author} {\bibfnamefont {G.}~\bibnamefont {Harder}}, \ and\ \bibinfo {author} {\bibfnamefont {C.}~\bibnamefont {Silberhorn}},\ }\href {\doibase 10.1103/PhysRevLett.120.213601} {\bibfield  {journal} {\bibinfo  {journal} {Phys. Rev. Lett.}\ }\textbf {\bibinfo {volume} {120}},\ \bibinfo {pages} {213601} (\bibinfo {year} {2018})}\BibitemShut {NoStop}%
\bibitem [{\citenamefont {Yan}\ \emph {et~al.}(2024)\citenamefont {Yan}, \citenamefont {Li}, \citenamefont {Hou}, \citenamefont {Zhu}, \citenamefont {Xiang}, \citenamefont {Li},\ and\ \citenamefont {Guo}}]{yan2024}%
  \BibitemOpen
  \bibfield  {author} {\bibinfo {author} {\bibfnamefont {W.-Z.}\ \bibnamefont {Yan}}, \bibinfo {author} {\bibfnamefont {Y.}~\bibnamefont {Li}}, \bibinfo {author} {\bibfnamefont {Z.}~\bibnamefont {Hou}}, \bibinfo {author} {\bibfnamefont {H.}~\bibnamefont {Zhu}}, \bibinfo {author} {\bibfnamefont {G.-Y.}\ \bibnamefont {Xiang}}, \bibinfo {author} {\bibfnamefont {C.-F.}\ \bibnamefont {Li}}, \ and\ \bibinfo {author} {\bibfnamefont {G.-C.}\ \bibnamefont {Guo}},\ }\href {\doibase 10.1103/PhysRevLett.132.080202} {\bibfield  {journal} {\bibinfo  {journal} {Phys. Rev. Lett.}\ }\textbf {\bibinfo {volume} {132}},\ \bibinfo {pages} {080202} (\bibinfo {year} {2024})}\BibitemShut {NoStop}%
\bibitem [{\citenamefont {De}\ \emph {et~al.}(2024)\citenamefont {De}, \citenamefont {Ansari}, \citenamefont {Sperling}, \citenamefont {Barkhofen}, \citenamefont {Brecht},\ and\ \citenamefont {Silberhorn}}]{de24}%
  \BibitemOpen
  \bibfield  {author} {\bibinfo {author} {\bibfnamefont {S.}~\bibnamefont {De}}, \bibinfo {author} {\bibfnamefont {V.}~\bibnamefont {Ansari}}, \bibinfo {author} {\bibfnamefont {J.}~\bibnamefont {Sperling}}, \bibinfo {author} {\bibfnamefont {S.}~\bibnamefont {Barkhofen}}, \bibinfo {author} {\bibfnamefont {B.}~\bibnamefont {Brecht}}, \ and\ \bibinfo {author} {\bibfnamefont {C.}~\bibnamefont {Silberhorn}},\ }\href {\doibase 10.1103/PhysRevResearch.6.L022040} {\bibfield  {journal} {\bibinfo  {journal} {Phys. Rev. Res.}\ }\textbf {\bibinfo {volume} {6}},\ \bibinfo {pages} {L022040} (\bibinfo {year} {2024})}\BibitemShut {NoStop}%
\bibitem [{\citenamefont {Robertson}(1929)}]{robertson1929}%
  \BibitemOpen
  \bibfield  {author} {\bibinfo {author} {\bibfnamefont {H.~P.}\ \bibnamefont {Robertson}},\ }\href@noop {} {\bibfield  {journal} {\bibinfo  {journal} {Phys. Rev.}\ }\textbf {\bibinfo {volume} {34}},\ \bibinfo {pages} {163} (\bibinfo {year} {1929})}\BibitemShut {NoStop}%
\end{thebibliography}%

\end{document}